\documentstyle[11pt]{article}
\input FEYNMAN
\title{\vspace{-1.in} \hfill {\small\rm TUM-HEP-341/99, SFB-375/340} \\~\\~\\
Neutrinos, their Partners and Unification}
\author{ George Triantaphyllou\thanks{e-mail:georg$@$ph.tum.de}
$\;$\\~  \\{\it Institut f\"ur Theoretische Physik, Technische 
Universit\"at M\"unchen}\\
{\it James-Franck-Strasse, D-85748 Garching, GERMANY }}
\begin{document}
\setlength{\baselineskip}{24pt}
\maketitle
\thispagestyle{empty}
\begin{abstract}
Efforts to unify group-theoretically
the standard-model gauge interactions with
the generation structure of fermions and their mirror
partners should be
accompanied with the unification of the corresponding gauge
couplings. In this paper the possibility of such a unification
is studied and conclusions on possible symmetry-breaking channels and scales, 
as well as on the fermion content of the theory, are drawn.
The breaking of some
of the symmetries allows various Majorana masses for neutrinos
and their mirror partners, so these are studied next.
Implications to neutrino mixings and 
mass hierarchies in connection with recent experimental results, 
as well as to electroweak precision tests, are then
discussed.
\vspace{2.in}
\end{abstract}
\vfill
\setcounter{page}{0}
\pagebreak

\section{Introduction}

The quantum numbers  of the known fermions under the  standard-model
gauge structure  allow their partial classification under 
the fundamental representations of the corresponding symmetry groups. 
This motivates efforts to complete this classification by studying
unifying symmetry groups large enough to 
accommodate all the fundamental-particle generations which have been observed
so far. Apart from the purely theoretical interest in such a possibility,
quests for such a unification usually lead to predictions on the existence of
new particles, like extra fermions and
gauge bosons \cite{georg}-\cite{Bars}. 
In particular, the new fermions are usually referred to as
the ``mirror partners" of the standard-model fermions.

Since there are currently several theoretical
and possibly also some experimental indications hinting for the existence of 
physics beyond the standard model at scales on 
the order of 1 TeV \cite{georg},
it is worthwhile investigating whether extensions of particle theories
in a direction compatible with generation 
unification could be related to these
indirect indications at TeV-energy scales.
Furthermore, in view of the fact that accelerators designed to operate
in the next decade plan to cover such high energies, it is quite
important to investigate their discovery potential by producing 
directly the particles predicted by the aforementioned extensions.

Before embarking in such a detailed production-and-decay study 
however, one should first check the internal consistency of the 
proposed theories and their compatibility with current experimental
constrains. A first effort to reproduce the observed charged-fermion 
mass hierarchy, the quark mixing matrix elements and the weak scale
while staying in agreement with the electro-weak precision data  within
such a framework 
was recently presented \cite{georg}. The purpose of the present work is
to tackle some related, equally important open issues.  

One of these issues is
to calculate the evolution of the gauge couplings to very high energies, 
in order to see if there is a sequence of symmetry breakings 
consistent with the unification picture which motivated the proposed 
extension in the first place. In all cases discussed,
the symmetries in question are taken to break spontaneously, and  getting
into the details of the breaking mechanism, like it being of dynamical
or fundamental nature and the transformation properties of its
non-zero vacuum-expectation-value, is avoided because this usually  
involves a high degree of arbitrariness and speculation in an area of no  
phenomenological input.  

Since the energy scales of these
breakings could be associated with the
lightness of the standard-model
neutrinos via the see-saw mechanism, the question of
neutrino masses and mixings left open in \cite{georg}
has to be studied next. This 
also allows the calculation of novel ``oblique"
contributions to the electro-weak 
parameters due to the possible Majorana nature of the mirror 
neutrinos.
An effort to address these different but closely related
issues follows next. 

\section{Coupling unification}

\subsection{Preliminary considerations}

The starting point of the discussion could be either of the unification 
gauge groups
$E_{8_{1}} \times E_{8_{2}}$ or $SO(16)_{1} \times SO(16)_{2}$ without
change in  the final results, with
gauge couplings $g_{1}$ and $g_{2}$ corresponding to the
groups with subscripts 1 and 2 respectively. These symmetries 
are taken to break at the unification scale $\Lambda_{GUT}$ down to 
$SO(10)_{1} \times SU(4)_{1G} \times SO(10)_{2} \times SU(4)_{2G}$.   
The fermions and mirror fermions of interest transform under the above
groups like ${\bf (16,\; {\bar 4},\; 1,\; 1)}$ and 
${\bf (1,\; 1, \; {\bar 16},\; 4)}$ respectively. 

It is  imagined next that around the same unification 
scale, $SO(10)_{1}\times 
SO(10)_{2}$ breaks to its diagonal subgroup
$SO(10)_{D}$, which has  accordingly a  gauge coupling at
that scale equal to 
\begin{equation}
g=\frac{g_{1}g_{2}}{\sqrt{g_{1}^{2}+g_{2}^{2}}}. 
\end{equation} 
This in its turn is taken to
break again at $\Lambda_{GUT}$ to its maximal subgroup
$SU(4)_{PS} \times SU(2)_{L} \times SU(2)_{R}$. 
In addition, 
$SU(4)_{1G} \times SU(4)_{2G}$ could be taken to break down at the same 
scale to either 
of the three groups $SU(3)_{2G}$,  $SU(3)_{2G} \times U(1)_{G}$, 
or  $SU(4)_{2G}$. 
Under the resulting group 
$SU(4)_{PS} \times SU(2)_{L} \times SU(2)_{R} \times SU(3)_{2G}$ for
instance, the standard-model-type fermions transform like four copies 
(``generations") of
${\bf (4,\;2,\;1,\;1)}$ and ${\bf ({\bar 4},\;1,\;2,\;1)}$, and the
mirror fermions like 
${\bf (4,\;1,\;2,\;3})+{\bf (4,\;1,\;2,\;1})$ 
and ${\bf ({\bar 4},\;2,\;1,\;3})+{\bf ({\bar 4},\;2,\;1,\;1})$. 
The two other possibilities will be discussed in subsection 2.4.
In all these cases it is assumed that the fourth generation 
standard-model-type fermions  
 pair-up with their mirror partners, acquiring   
gauge-invariant masses on the order of the $SU(4)_{2G}$ breaking
scale. (This is $\Lambda_{GUT}$ for the first two cases and about
1 TeV for the third case, as will be seen in subsection 2.4.) 

One first notes that the only way to unify the generation-group
coupling with the other gauge couplings is to satisfy the relation $g_{1} \gg 
g_{2}$ at $\Lambda_{GUT}$, 
because then the common unification coupling is $g \approx g_{2}$
according to Eq. 1.
Therefore, the generation group $SU(4)_{1G}$, which  
is taken to break completely at $\Lambda_{GUT}$,  
is strongly coupled at that scale. 
The situation  with generation group $SU(3)_{2G}$ with coupling
$g_{2}$ is first investigated. 
The basis of the analysis of the gauge-coupling renormalization
that  follows 
is more of a qualitative nature and limited to the one-loop 
$\beta$-function, due to 
the many uncertainties of the dynamics influencing the running of these  
couplings. 
These are mainly due not only to our 
ignorance of the exact masses of the mirror fermions
and of the type of new physics needed to break the gauge groups 
involved in this picture, which
could be Higgs particles in various presently unpredictable
representations, but also to the  possible 
existence of supersymmetric partners of the standard-model fermions and
to threshold effects near the unification scale.

These uncertainties  lead one to take all the mirror fermions to have the
same mass $\Lambda_{M}$ at around 1 TeV
for simplicity, since  the coupling unification
is found to be anyway quite 
insensitive to this scale. 
Below $\Lambda_{M}$ the couplings evolve like in the standard model.
Above that scale, one has to take the mirror fermions into account.
It is  then assumed that there exists a
``desert" between $\Lambda_{M}$ and the Pati-Salam scale $\Lambda_{PS}$
where $SU(4)_{PS}$ is broken,
with no new dynamics or particles able to influence the evolution of the
gauge couplings with energy. 

The $\beta$-function describing the evolution of the gauge coupling $g$ of
an $SU(N)$ group with $N_{f}$ fermion
$N$-plets with respect to momentum $p$ is given by
\begin{equation}
\beta \equiv \frac{dg}{d\ln{(p/p_{0})}} 
= - \frac{g^{3}}{48\pi^{2}}({11N}-2N_{f}+r) 
\end{equation}

\noindent where $r$ stands for higher-than-one-loop 
corrections and $p_{0}$ is some reference scale. 
If the same fermions
transform also under the fundamental representation of  another unitary 
gauge group $SU(N^{\prime})$ with 
coupling $g^{\prime}$ much larger than $g$, the quantity $r$ at two
loops is approximately given by 
\cite{Jones} 
\begin{equation}
r=\frac{g^{\prime \; 2}(N^{\prime \; 2}-1)}
{32 \pi^{2}}.
\label{eq:loop}
\end{equation} 

Therefore, when the $SU(3)_{2G}$ interactions become strong at around
2 TeV and break $SU(2)_{L}\times U(1)_{Y}$ dynamically by an
effective Higgs mechanism induced by fermion condensates   
\cite{georg}, the corresponding fine-structure-constant is 
$\alpha_{G} \approx 1$. One therefore gets
$r \approx 0.3$, which is still much smaller than
the one-loop contribution to the other couplings, even for the smaller
groups considered like $SU(2)_{L}$ or $U(1)_{Y}$ (the influence of 
the other couplings to each other is of course even more negligible due
to their smallness). In addition, since
$SU(3)_{2G}$ is taken to break just after it becomes strong \cite{georg}, 
it has a rather limited energy region where it can influence
substantially the $\beta$ functions of interest, so  
large  deviations from the one-loop renormalization of 
the rest of the gauge
couplings are not expected. This issue is investigated further by
presenting a particular example in subsection 2.3. 

Moreover,  
a fundamental Higgs mechanism for breaking $SU(3)_{2G}$ is avoided  by
evoking the mechanism conjectured in the Appendix of \cite{georg}. In 
any case, a minimal fundamental 
Higgs mechanism breaking the generation symmetry, apart from  all
the naturalness problems it carries with it, 
would make the corresponding gauge coupling run slightly slower.
The generation-coupling
unification with the rest of the gauge couplings at $\Lambda_{GUT}$ 
would then still be achievable by slightly lowering the maximal value
this coupling reaches before $SU(3)_{2G}$  breaks and/or lowering the
mirror-mass scale $\Lambda_{M}$. An effort to estimate the energy scales
entering this problem without fundamental scalars is presented in the
next subsection. 

However simple, the approach adopted 
 allows  to draw general conclusions, not depending on 
particular details, on the way the unification
groups break down to the standard-model gauge structure. It has to
be stressed nevertheless that the class of symmetry-breaking channels of  
interest here has an additional degree of freedom compared to the usual 
or to the 
supersymmetric unification, and that is the Pati-Salam symmetry
breaking scale $\Lambda_{PS}$. This can be in most cases 
slightly adjusted to
 allow unification of couplings even after the correct
inclusion of these corrections, unless one introduces
unnaturally large Higgs sectors to break the gauge symmetries. 
The results that follow should therefore
be seen not as exact predictions but rather as order-of-magnitude estimates. 

\subsection{Calculation of $\Lambda_{GUT}$ and proton life-time}

The analysis presented here is based on different alternative 
breakings of the  gauge symmetry $SU(4)_{PS}$, since {\it a priori}
there is no obvious reason to expect a specific breaking channel.  
The subsequent analysis will show that only one alternative seems to
be viable if one takes proton-lifetime bounds and the order of magnitude
of the weak scale into consideration. It is particularly
interesting therefore to note that, under certain assumptions,
current phenomenological input is able
to constrain the number of different group-breaking channels, even when
these appear at scales
 much higher than the ones directly accessible at present. 

In particular, 
the Pati-Salam group is  taken to break at the scale $\Lambda_{PS}$
either along the channel
\begin{equation}
SU(4)_{PS} \times SU(2)_{R} \longrightarrow SU(3)_{C} \times U(1)_{Y}, 
\end{equation}
or along the channel 
\begin{equation}
SU(4)_{PS} \times U(1)_{R} \longrightarrow SU(3)_{C}
\times U(1)_{Y}
\end{equation}
 if the breaking $SU(2)_{R} \longrightarrow U(1)_{R}$ has 
already occurred at $\Lambda_{GUT}$. A third possibility is also
examined, namely one of a Pati-Salam symmetry breaking like 
\begin{equation}
SU(4)_{PS} \longrightarrow
SU(3)_{C} \times U(1)_{B-L}
\end{equation}
at $\Lambda_{GUT}$, which is followed by the breaking of 
$SU(2)_{R}\times U(1)_{B-L} \longrightarrow U(1)_{Y}$ at scale $\Lambda_{R}$. 
In all these alternative scenarios, the Pati-Salam 4-plets are each broken to
a QCD triplet and a lepton, while simultaneously giving rise to 
a ``predecessor" of the electromagnetic charge. 

It is also noted that, as a first approximation, 
below the mirror-mass threshold scale $\Lambda_{M}$ the couplings of 
all the non-abelian groups except for the generation group
are taken to evolve with $N_{f}=6$ like in the standard model, 
and above $\Lambda_{M}$ with 
$N_{f}=12$, the doubling being caused by the
the existence of mirror fermions which  leads to an 
abrupt change in slope to the  
running of the couplings at that scale. The eventual top-quark decoupling and
the mixing between ordinary and mirror fermions, which
apart from the top-quark is quite small, is thus also neglected. 
The $SU(3)_{2G}$ coupling evolves at all scales with 
$N_{f}=8$. These $N_{f}$ values are the same for all three Pati-Salam 
breaking channels considered.

It is more convenient to work in the following
with the inverse structure constants $\alpha^{-1}=4\pi/g^{2}$
since their evolution is linear with $\ln{(p/p_{0})}$. 
The value of the hypercharge coupling 
$\alpha_{Y}(\Lambda_{PS})$ in the first two cases is computed via the relation 
\begin{equation}
\alpha^{-1}_{Y} = (3\alpha^{-1}_{R}+2\alpha^{-1}_{PS})/5
\end{equation}
which is evaluated
at the Pati-Salam scale $\Lambda_{PS}$, where
$\alpha_{R}$ is the coupling corresponding to $SU(2)_{R}$ or $U(1)_{R}$
respectively. 
In the third case   the
hypercharge coupling is given by the relation 
\begin{equation}
\alpha^{-1}_{Y}(\Lambda_{R}) = (3\alpha^{-1}_{R}(\Lambda_{R})
+2\alpha^{-1}_{B-L}(\Lambda_{R}))/5. 
\end{equation}
Furthermore, the first and third 
cases are based on the working assumption of unbroken
discrete left-right symmetry above the scale where $SU(2)_{R}$ is 
broken, i.e.  $\alpha_{R}=\alpha_{L}$.
As was said in the introduction, a discussion on the possible 
breaking mechanisms of these symmetries is here avoided, since the purpose
of the analysis is to allow 
general qualitative conclusions to be drawn.

\begin{figure}[t]
\vspace{4.1in}
\includegraphics{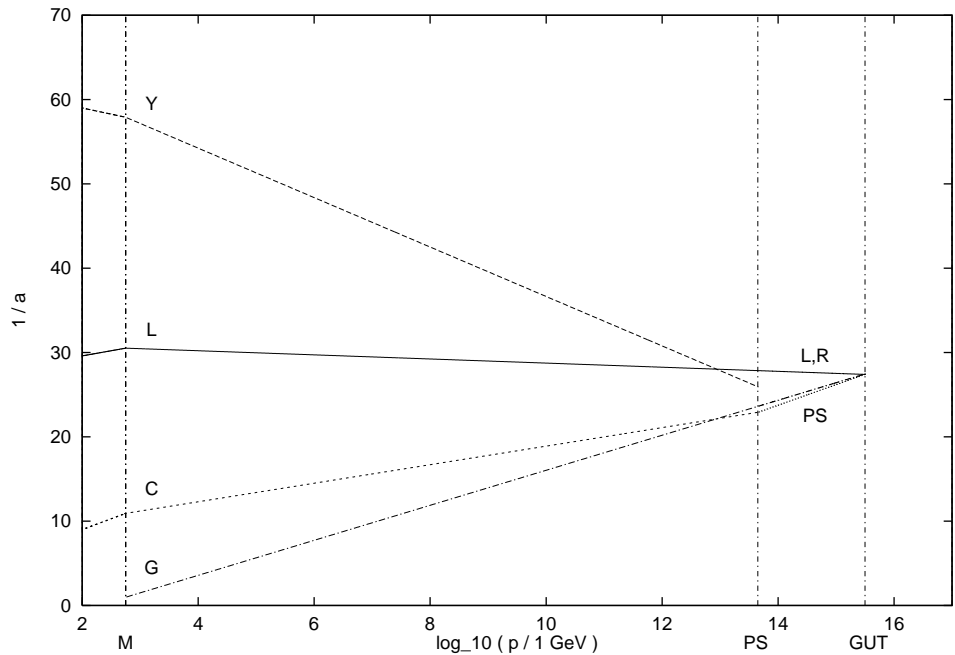}
\caption{The running  of the inverse fine-structure constants
$\alpha^{-1}_{Y,L,C,G}$
and later $\alpha^{-1}_{R,PS}$,  
corresponding to the breaking channel
$SU(4)_{PS} \times SU(2)_{R} \longrightarrow SU(3)_{C} \times U(1)_{Y}$.
The vertical lines, starting from small
energies, correspond to the
scales $\Lambda_{M}$, $\Lambda_{PS}$ and $\Lambda_{GUT}$. 
The relevant scales are found to be
$\Lambda_{M}=10^{2.75}$ GeV,   
$\Lambda_{PS}=10^{13.65}$ GeV and $\Lambda_{GUT}=10^{15.5}$ GeV.} 
~\\
\label{fig:fig1}
\end{figure}

\begin{figure}[t]
\vspace{4.1in}
\includegraphics{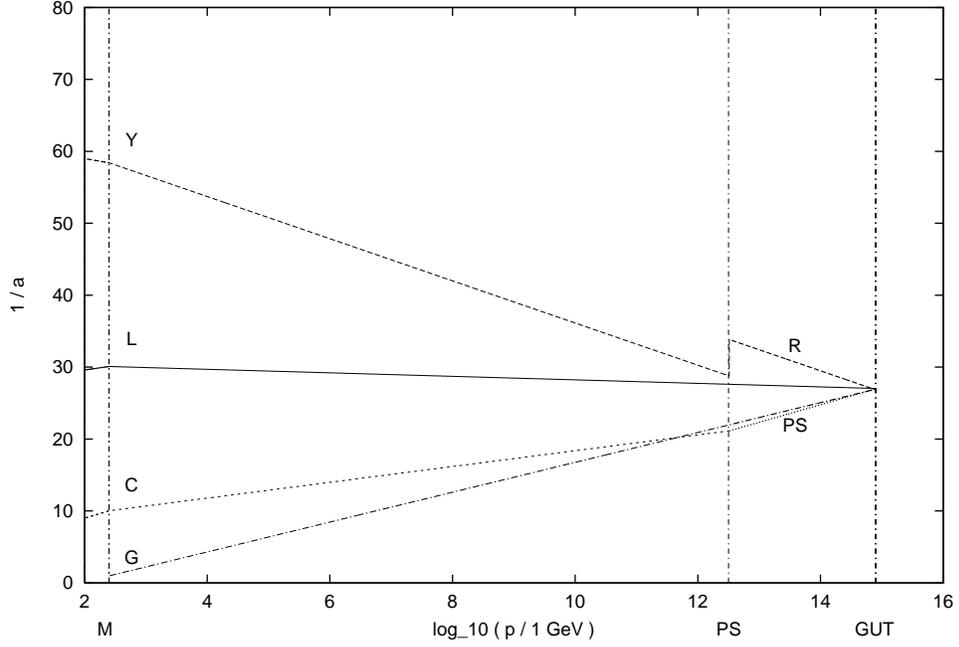}
\caption{The running  of the inverse fine-structure constants
$\alpha^{-1}_{Y,L,C,G}$
and later $\alpha^{-1}_{R,PS}$, corresponding to the breaking channel 
$SU(2)_{R} \longrightarrow U(1)_{R}$ at $\Lambda_{GUT}$ and 
$SU(4)_{PS} \times U(1)_{R} \longrightarrow SU(3)_{C} \times U(1)_{Y}$
at $\Lambda_{PS}$.
The scales are
$\Lambda_{M}=10^{2.4}$ GeV,   
$\Lambda_{PS}=10^{12.5}$ GeV, and $\Lambda_{GUT}=10^{14.9}$ GeV.} 
~\\
\label{fig:fig1a}
\end{figure}

\begin{figure}[t]
\vspace{3.5in}
\includegraphics{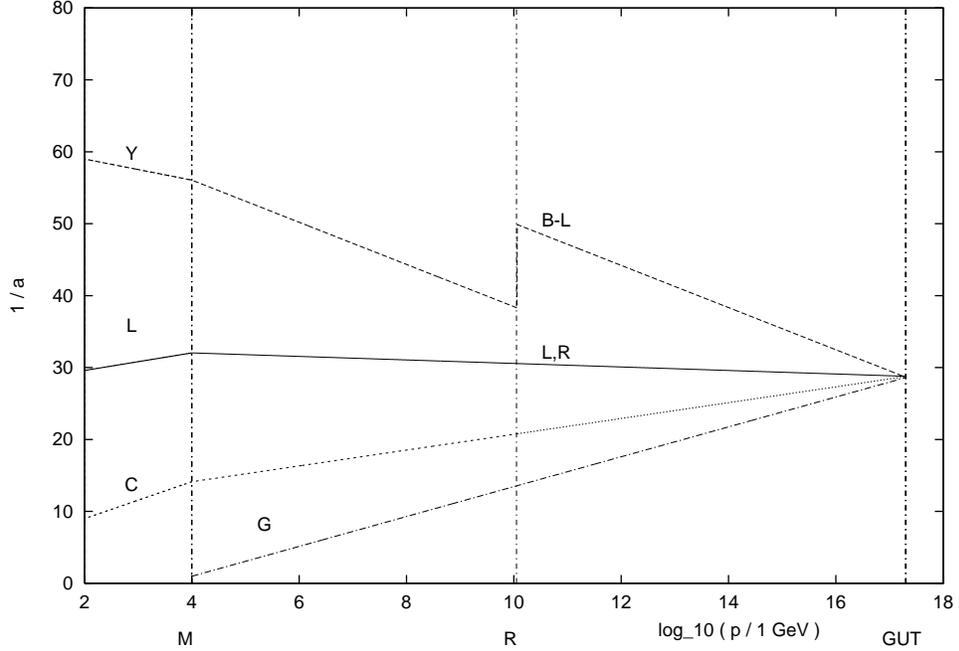}
\caption{The running  of the inverse fine-structure constants
$\alpha^{-1}_{Y,L,C,G}$
and later $\alpha^{-1}_{B-L,R}$, assuming that there is a symmetry
breaking channel like
$SU(4)_{PS} \longrightarrow SU(3)_{C} \times U(1)_{B-L}$ at
$\Lambda_{GUT}=10^{17.3}$ GeV and  
$SU(2)_{R} \times U(1)_{B-L} \longrightarrow U(1)_{Y}$
at $\Lambda_{R}=10^{10.05}$ GeV. 
The vertical lines, starting from small
energies, correspond here to the
scales $\Lambda_{M}$, $\Lambda_{R}$ and $\Lambda_{GUT}$. 
The mirror-fermion masses are
taken to be  $\Lambda_{M}=10^{4}$ GeV in order to allow the generation
coupling to meet the rest of the gauge couplings.}  
~\\
\label{fig:fig4}
\end{figure}

The starting point of the calculation is based on  the   
approximate experimental values for these quantities listed 
below \cite{DataBook} 
\begin{eqnarray}
\alpha^{-1}_{Y}(M_{Z})&\sim&59.2, \nonumber \\ 
\alpha^{-1}_{L}(M_{Z})&\sim&29.6,  \nonumber \\ 
\alpha^{-1}_{C}(M_{Z})&\sim&8.4. 
\end{eqnarray}
Moreover, at scale $\Lambda_{M}$ 
the $SU(3)_{2G}$ coupling is taken to be  
equal to $\alpha_{2G}(\Lambda_{M})=1$.  
This coupling is not plotted for 
$\alpha_{2G} >1$, because then higher-order corrections to the
renormalization of this coupling become important. This  is
expected to have a limited effect to the other
couplings however, since  the generation group
breaks at around the same scale. This issue is examined again later.  

The evolution of the inverse fine-structure constants
$\alpha^{-1}_{i}$ for the various couplings $i=Y,L,C,G,PS,R$
of the groups $U(1)_{Y}$, $SU(2)_{L}$, $SU(3)_{C}$, $SU(3)_{2G}$, 
$SU(4)_{PS}$ and $SU(2)_{R}$
corresponding to the three different breaking channels mentioned above
are plotted consecutively in Figures 1, 2 and 3. 
The relevant scales for which unification is possible, along with 
the value of the unification coupling are also given in Table 1. 

   \begin{table}[t]
    \begin{tabular}
    {||c||
    c@{\hspace{2mm}}|@{\hspace{2mm}}c@{\hspace{2mm}}
 |@{\hspace{2mm}}c@{\hspace{2mm}}|@{\hspace{2mm}}c
 ||c@{\hspace{2mm}}||} \hline \hline
\rule[-3mm]{0cm}{8mm}
Symmetry breaking sequence
&\multicolumn{4}{c||}{Energy scales (GeV)} &  
  \\  \cline{2-5} 
\rule[-3mm]{0cm}{8mm}
assuming an $SU(3)_{2G}$ generation group 
&$\Lambda_{M}$
&$\Lambda_{R}$
&$\Lambda_{PS}$
&$\Lambda_{GUT}$&\raisebox{2.3ex}[-2.3ex]{$\alpha_{GUT}$}
 \\ \cline{1-6} \hline \hline 
\rule[-3mm]{0cm}{8mm}$SU(4)_{PS}\times SU(2)_{R} 
\rightarrow SU(3)_{C}\times U(1)_{Y}$&$10^{2.75}$&$\Lambda_{PS}$
&$10^{13.65}$&$10^{15.5}$&0.036  
\\ \cline{1-6}
\rule[-3mm]{0cm}{8mm}$SU(4)_{PS}\times U(1)_{R}
\rightarrow SU(3)_{C}\times U(1)_{Y}$&$10^{2.4}$&$\Lambda_{PS}$
&$10^{12.5}$&$10^{14.9}$&0.037  
\\ \cline{1-6}
\rule[-3mm]{0cm}{8mm}$SU(4)_{PS}\rightarrow SU(3)_{C}\times U(1)_{B-L}$
&$10^{4}$&$10^{10.05}$&$\Lambda_{GUT}$&$10^{17.3}$&0.034 
\\ \hline \hline 
      \end{tabular}  
 \caption{ The energy scales  required to achieve unification assuming
 three different symmetry-breaking channels, and the corresponding
 value of the unification coupling. The second channel is disfavored
 because of the low unification scale, and the third channel is also
 disfavored because of the large mirror-fermion masses implied by
 $\Lambda_{M}$.  
}  
~\\
     \end{table}

To begin with, we discuss the first two cases. The energy
scales of interest are found to be
$\Lambda_{M} = 10^{2.75}$ GeV and $\Lambda_{PS} = 10^{13.65}$ GeV 
in the first case, and  
$\Lambda_{M} = 10^{2.4}$ GeV and $\Lambda_{PS} = 10^{12.5}$ GeV in the
second. The corresponding unification scales and couplings
are found to be $\Lambda_{GUT} = 10^{15.5}$ GeV and $\alpha_{GUT}=0.036$
in the first case,
and $\Lambda_{GUT} = 10^{14.9}$ GeV and $\alpha_{GUT}=0.037$
in the second. One can see  
at the scale $\Lambda_{PS}$ in Figs. 1 and 2 the characteristic
change in slope of the Pati-Salam coupling
 when the group $SU(4)_{PS}$ breaks down to $SU(3)_{C}$, which is due to
 the different quadratic Casimirs of their adjoint representations, as well
as the starting of the hypercharge-coupling running at that scale, since
at scales higher than $\Lambda_{PS}$, $U(1)_{Y}$ is embedded in other groups. 

The inclusion of a minimal Higgs field able to break these symmetries
spontaneously would for the same $\Lambda_{GUT}$
slightly shift $\Lambda_{PS}$ downwards since it 
would slow down somewhat the running of the Pati-Salam coupling. 
It is clear from the figures that, with the
present fermion content, the slopes of the gauge couplings
do not favor $SU(5)$ unification. Also, the slope of  $SU(2)_{L}$ below
$\Lambda_{M}$ speaks against the addition of new weak-singlet fermions, as
is usually done in universal see-saw models \cite{Cho}, if one seeks
coupling unification. 

The third alternative breaking sequence is as already said
to break the Pati-Salam group
at the unification group like $SU(4)_{PS} \longrightarrow SU(3)_{C} \times 
U(1)_{B-L}$, and have later the breaking $SU(2)_{R} \times U(1)_{R} 
\longrightarrow U(1)_{Y}$ at scale $\Lambda_{R}$. 
This possibility is drawn in Figure 3. The relevant
scales are found to be $\Lambda_{M}  = 10^{4}$ GeV, 
$\Lambda_{R} = 10^{10.05}$ GeV and $\Lambda_{GUT} = 10^{17.3}$ GeV.   
The scale $\Lambda_{R}$ with the present fermion content is quite
large (compare for instance solutions with alternative fermion
contents \cite{Man}), supporting a see-saw mechanism for the neutrino masses. 
The unification scale is in this case quite large, a result  
reminiscent of \cite{Deshp}, and the
common coupling at that scale is $\alpha_{GUT}=0.034$.
The main reason for the largeness of $\Lambda_{M}$ is the effort to 
unify the generation-group gauge coupling with the other couplings.

If this unification  condition is relaxed, in the same
way it is relaxed in connection with a $SU(4)_{2G}$ generation
group that is discussed in the next subsection, 
the rest of the couplings can be
unified with a smaller $\Lambda_{M}$ and this channel is still viable.  
In the present case nonetheless,  the  largeness of $\Lambda_{M}$ used 
would correspond to an unacceptably large
weak scale. The fact that $SU(2)_{R}$ breaks far away
from $\Lambda_{GUT}$ would also render  effects coming from the
mechanism to which its breaking is due and which are here
neglected, like the existence of scalar particles,  more important.
Such effects however are not expected to  alleviate the problem of 
the large scale $\Lambda_{M}$. We can therefore conclude here that
the third breaking channel is improbable, unless generation-coupling 
unification with the rest of the couplings is abandoned. Another way to keep 
$\Lambda_{M}$ small would be of course
to add a large Higgs sector transforming
non-trivially under $SU(3)_{2G}$, but this alternative 
is not investigated since it is foreign to  the 
present conceptual framework and would raise naturalness problems. 

Gauge-coupling unification in connection with bounds on the proton 
life-time is discussed next, since the breaking of $SO(10)_{D}$ at
$\Lambda_{GUT}$ can 
induce proton decay via effective four-fermion operators.
This issue could actually help us decide between the first two
breaking channels proposed. 
From the proton life-time experimental 
constraint 
\cite{DataBook} 
\begin{equation}
\tau (p \rightarrow e^{+}\pi^{0})> 5.5 \times
10^{32} \;{ \rm yr} 
\end{equation}
and the theoretical order-of-magnitude estimate 
\begin{equation}
\tau^{-1} \approx 
\alpha^{2}_{GUT}\frac{m_{p}^{5}}{\Lambda_{GUT}^{4}},  
\end{equation} one gets the inequality 
\begin{equation}
\alpha_{GUT} < 0.074 \left(\frac{\Lambda_{GUT}}{10^{15.5}} \right)^2.
\end{equation}

This proton life-time bound makes  clear that  the
second breaking channel
possibility is disfavored due to a small unification scale. 
Nevertheless, it cannot be at this point definitely excluded due to
the limited level of accuracy of the current rather qualitative analysis. 
Note that this result is reminiscent of the result of \cite{Deshp} in 
an analysis with the
same breaking sequence in a left-right symmetric context but without
mirror fermions. One is consequently left with the first alternative
as the one corresponding to the most probable symmetry breaking sequence. 

\begin{figure}[t]
\vspace{3.5in}
\includegraphics{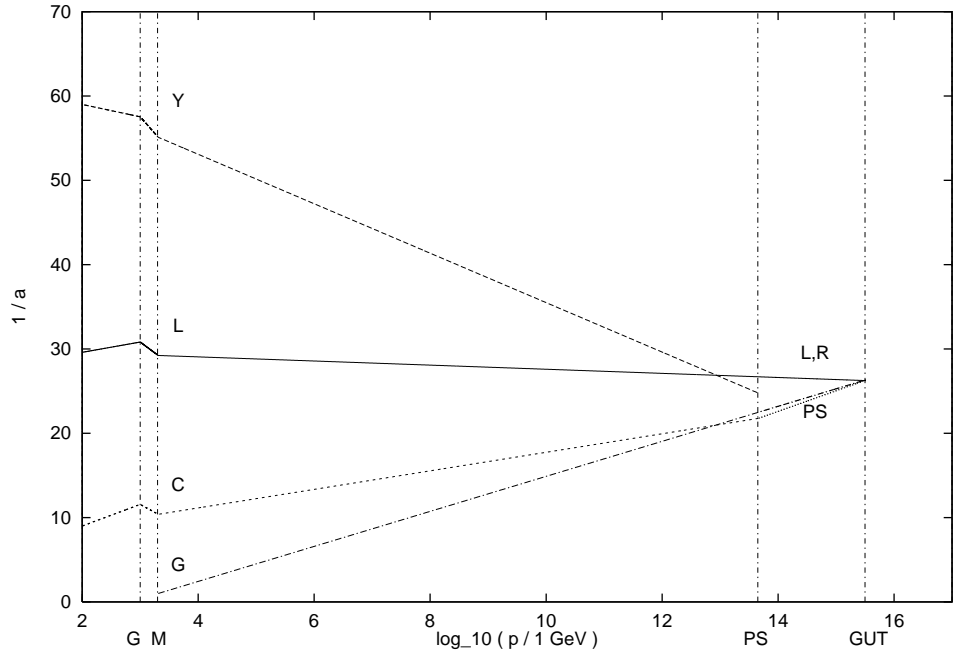}
\caption{The running of the inverse fine-structure constants
$\alpha^{-1}_{Y,L,C,G}$
and later $\alpha^{-1}_{R,PS}$,  
corresponding to the breaking channel
$SU(4)_{PS} \times SU(2)_{R} \longrightarrow SU(3)_{C} \times U(1)_{Y}$,
in an effort to simulate a possible
large influence of the strong generation group $SU(3)_{2G}$. 
It is found that in order to achieve unification one needs 
$\Lambda_{G}=10^{3}$ GeV,   
$\Lambda_{M}=10^{3.3}$ GeV, 
$\Lambda_{PS}=10^{13.65}$ GeV, and 
$\Lambda_{GUT}=10^{15.5}$ GeV.}   
~\\
\label{fig:fig2}
\end{figure}

\subsection{The effect of strong dynamics} 
Strong dynamics can alter the results quoted above, 
since higher-order corrections
to the various $\beta$-functions due to the strong $SU(3)_{2G}$ interactions
could become important if the quantity $r$ introduced before is not 
negligible.  However, as also noted in \cite{Holdom}, the  
fermion content of the theory implies that this effect, 
however large, would be  uniform for
all standard-model couplings, as is shown in Figure 4 for an 
exaggerated effect
corresponding to $r=40$. From Eq.\ref{eq:loop}, this would correspond to
a highly-non-perturbative
generation coupling $\alpha_{G} \approx 126$
(this number is of course purely indicative, since 
the perturbative $\beta$-function has no meaning in this 
regime), something which has the same 
influence on each of the other relatively weak $SU(N)$ couplings
as the introduction of 20 new fermion $N$-plets. 

Such strong dynamics can shift the unification
coupling $\alpha_{GUT}$ to larger values, but cannot shift
the unification scale.  In reality the coupling-evolution
curves shown should be smooth, without angles, but  
$r$ is here taken to become suddenly important 
for illustration purposes, in a - perhaps overambitious -  
effort to simulate the relevant effect. There is no
guarantee of course that the non-perturbative effects of $SU(3)_{2G}$ can be 
limited even by such  large $r$-values, 
but it is assumed that they are in order
to maintain the conclusions presented in this work. 

The analysis of this alternative leads one to  split
the scale where the mirror fermions decouple $\Lambda_{M}$, 
from the scale $\Lambda_{G}$ 
where the generation group becomes strong, and consider for instance 
 the most probable breaking channel corresponding to 
Fig. 1. 
An effort is therefore made to ``parametrize" by means of the
quantity $r$ our ignorance of the 
the strong dynamics, and the effects they have on the other couplings, in  
the energy region between the scales $\Lambda_{G}$ and $\Lambda_{M}$. 
The unification
and Pati-Salam scale remain the same as in Fig. 1, 
but the scale $\Lambda_{M}$ has to be
raised to $10^{3.3}$ GeV and one has  further 
 a new scale $\Lambda_{G}=10^{3}$ GeV  
in order to achieve unification. The strong
coupling tends to make the other couplings slightly
larger at $\Lambda_{GUT}$, i.e. one gets $\alpha_{GUT}=0.037$.
If these effects are really large, difficulties  with the
reproduction of the weak scale could potentially arise due to the heaviness
of the mirror fermions.

\begin{figure}[t]
\vspace{3.4in}
\includegraphics{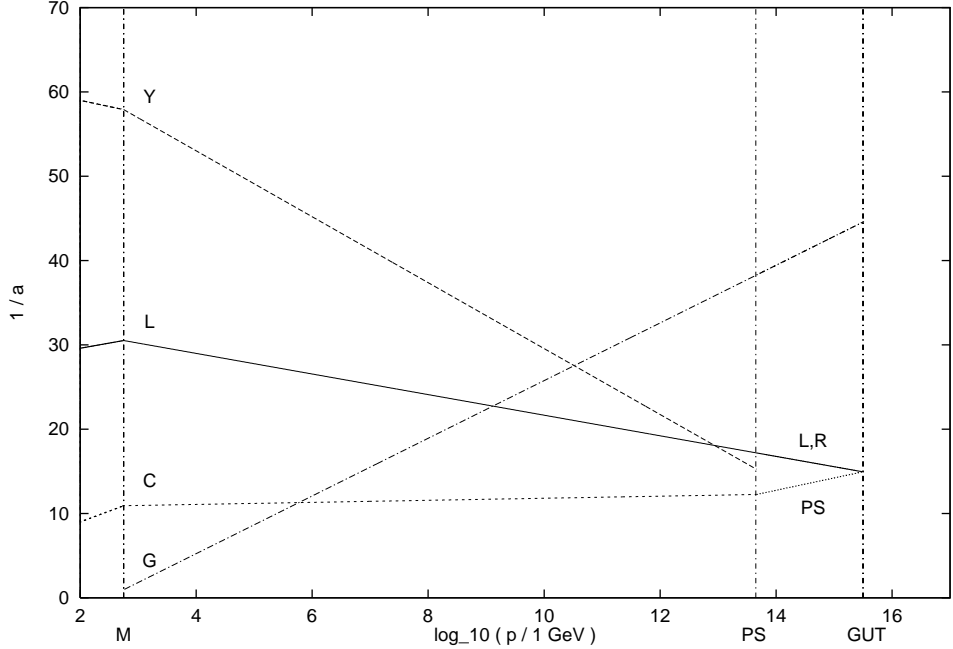}
\caption{The running of the inverse fine-structure constants
$\alpha^{-1}_{Y,L,C,G}$
and later $\alpha^{-1}_{R,PS}$,  
corresponding to the breaking channel
$SU(4)_{PS} \times SU(2)_{R} \longrightarrow SU(3)_{C} \times U(1)_{Y}$.  
The generation group $SU(4)_{2G}$ is here taken to be  
unbroken until TeV-scales, and unification of its coupling with the
rest of the gauge couplings is here
abandoned. All couplings apart from the generation
coupling are larger at the
unification scale due to 
the existence of two additional generations of 
fermions. The relevant scales are found to be as in Fig. 1
equal to 
$\Lambda_{M}=10^{2.75}$ GeV,    
$\Lambda_{PS}=10^{13.65}$ GeV, and 
$\Lambda_{GUT}=10^{15.5}$ GeV.} 
~\\
\label{fig:fig3}
\end{figure}

\subsection{Mirror-generation groups other than $SU(3)_{2G}$}

The issue of the generation groups comes next. 
Even if there is an abelian generation group $U(1)_{G}$ surviving down
to TeV-scales along with $SU(3)_{2G}$, unification requires 
that its coupling is negligibly small at 
low energies, so its running is neglected and 
its evolution with energy not plotted. Note however that if one wants
the see-saw
mechanism to work for the standard-model neutrinos, as will be
seen in the next section, this symmetry would
have to be broken at very large energy scales to allow for ultra-light
neutrinos. 
In the first case  considered for instance, one could  think of a breaking 
channel involving $U(1)_{G}$, like
$SU(4)_{PS} \times SU(2)_{R} \times U(1)_{G} \longrightarrow 
SU(3)_{C} \times U(1)_{Y} \times U(1)_{G^{\prime}}$,  
with the standard-model neutrinos being $U(1)_{G^{\prime}}$-neutral. 
This does not change the previous  
results and conclusions, but it could alter  
the values of the abelian generation group charges in \cite{georg}. 

One could also give-up unification of the generation group couplings
with the other couplings. This would mean either letting the relation
between $g_{1}$ and $g_{2}$ free, or considering $SU(4)_{2G}$ instead of
$SU(3)_{2G}$ unbroken
down to low-energy scales. The latter would correspond to having
also a fourth fermion generation paired-up with its mirror partner at scales
of the order of 1 TeV, which would make $N_{f}=16$ for the
standard-model groups instead of $N_{f}=12$
used in all previous cases. The generation group $\beta$-function
would remain with $N_{f}=8$ as before.
The corresponding running of the couplings 
is plotted in Figure 5. The scales $\Lambda_{M}$, $\Lambda_{PS}$ and
$\Lambda_{GUT}$ remain the same as in the more favored case of Fig. 1.

This scenario  has the advantage that it can generate lepton masses through
a strong $U(1)_{G}$ coupling after its breaking at around
1 TeV \cite{georg}.  
It suffers however from 
the same problem as the one encountered in \cite{Wil}, since in both
cases the generation coupling is running too fast and complete unification
is lost.   Unification could only be achieved by pushing the 
scale $\Lambda_{M}$ to very high values and thus  paying the
unacceptable price of a weak scale several orders of magnitude larger
than what it should be.
Moreover, due to the introduction of additional 
low-lying fermions, it pushes the other gauge couplings to larger values
at the unification point, since one finds $\alpha_{GUT}=0.067$,
something which could potentially create problems with proton decay. 
This is also the reason why no additional fermion generations and their
mirrors should be generally expected much below the unification scale. 

Important
conclusions can be therefore drawn here, namely that for 
unification of the generation coupling with the other gauge couplings 
one should have a group $SU(3)_{2G}$ becoming strong at around
1 TeV, whose coupling
renormalization naturally reproduces  the hierarchy between the QCD and
the weak scale, and which argues against alternative generation
groups unbroken to low energies like $SU(4)_{2G}$ or $SU(2)_{2G}$ for 
instance. This is to the best of our knowledge the first example of a 
fully unifiable and phenomenologically 
viable dynamical symmetry breaking model. 

Note moreover that in all cases considered, the
Pati-Salam group has to break at a very high energy to 
achieve unification, which is
due to the fast running of the Pati-Salam coupling,
making it very difficult
to feed-down quark masses to leptons. 
In order to avoid light fundamental Higgses therefore,  
the existence of gauge-invariant operators of the form
${\bar l_{R}}l_{L}^{M}{\bar q_{R}^{M}}q_{L}$ which arise
non-perturbatively would have to be postulated to generate lepton masses. 

\section{Neutrino masses and mixings}

\subsection{The structure of the mass matrix}

Having acquired a general idea on how and at what energy scales 
the gauge symmetries  break in this model, 
an investigation of the neutrino
mass hierarchy and mixing angles is presented next. 
Since the symmetry-breaking channels considered  involve always
the breaking of the $SU(2)_{R}$ and $SU(3)_{2G}$
symmetries which protect neutrino
and mirror neutrino Majorana masses respectively, it is most natural
to ask now what happens to their relevant masses and mixings after
these breakings.
In light of recent experimental 
results on neutrinos \cite{Bilenky}, 
such considerations go beyond  a mere academic 
interest. 

 Even though the analysis that follows is not
exhaustive and the mass-matrix entries do not  stem from a specific
calculational scheme, this example
not only demonstrates explicitly the power that the present framework
has to describe several phenomenological issues without
fine-tuning of parameters, but it also  
leads to quite useful general conclusions about the
structure of  the mirror-lepton submatrices. 
For simplicity, the lepton mass matrices, 
as well as their  
submatrices $M_{L}, m_{L}, M, {\tilde M}, m_{I}, m_{B}$
defined below, are  taken symmetric and real, 
ignoring $CP$ violating phases which may arise in the lepton sector. 

Recalling  the numerical example in \cite{georg}, for 
the charged leptons $l$
a $ 6\times 6$ mass matrix ${\cal M}_{L}$ is introduced,
having the form 
\begin{eqnarray}
 && 
\;\;\;\;\; l_{L}  
 \;\;\;\;\;\; l^{M}_{L} 
 \nonumber \\
 \begin{array}{c} 
 {\bar l}_{R} \\ 
 {\bar l}^{M}_{R}  
 \end{array} &&
\left(\begin{array}{cc}  0 & m_{L} \\ m_{L} & M_{L}
 \end{array} \right)  
 \end{eqnarray}

\noindent with the superscripts $M$ indicating mirror fields and
$M_{L}$, $m_{L}$ being $3 \times 3$ matrices in generation space
given by 
\begin{eqnarray}
\begin{array}{c}  \\ M_{L} ({\rm GeV}) =\\ \end{array} &&
\left(\begin{array}{ccc} 180 & 0 & 0 \\ 0 & 200 & 0 \\
0 & 0 & 200 \end{array} \right)
\begin{array}{c}  \\, \; m_{L} ({\rm GeV}) =\\ \end{array}
\left(\begin{array}{ccc} 0.25 & 0.25 & 0.1 \\ 0.25 & 3.8 & 1 \\
0.1 & 1 & 17 \end{array} \right).
\end{eqnarray}
 
\noindent These produce after diagonalization
the following lepton and mirror-lepton
mass hierarchy (at 2 TeV and in GeV units):

\noindent Standard-model charged leptons \hfill Mirror charged leptons $\;$ \\
$m_{\tau} =1.45$ , $m_{\mu}=0.07$,
$m_{e}=3\times10^{-4}$  \hfill
$m_{\tau^{M}} = 201$,
$m_{\mu^{M}} = 200$, $m_{e^{M}} = 180$.
Recall that the order of magnitude of these masses has to fulfill the double
requirement of escaping direct detection in current high-energy facilities
and reproducing along with the other mirror fermions the correct weak scale. 

For the neutrinos the situation is more complicated, since they could
be of a Majorana nature.
The neutrino $12 \times 12$ mass matrix ${\cal M}_{N}$ is
introduced next, and it is taken to  have the following form: 
\begin{eqnarray}
 && 
 \;\;\;\;\; \nu_{L} \;\;\;\;\; \nu^{c}_{R} 
 \;\;\; \nu_{L}^{M} \;\;\;\; \nu^{M\;c}_{R} 
 \nonumber \\
 \begin{array}{l} 
 {\bar \nu^{c}}_{L} \\ {\bar \nu}_{R} \\ 
 {\bar \nu^{M\;c}}_{L} \\ {\bar \nu^{M}}_{R} 
 \end{array} &&
 \left(\begin{array}{cccc} 
0      & 0      & 0          & m_{I}\\
0      & M      & m_{I}      & 0     \\
0      & m_{I}  & {\tilde M} & m_{B}     \\
m_{I}  & 0      & m_{B}      & 0 
 \end{array} \right),  
 \end{eqnarray}

\noindent where the entries shown are $3 \times 3$ matrices in 
generation space. The zero blocks are protected by the $SU(2)_{L}$ 
symmetry. The matrices $m_{B}$ and $m_{I}$ denote Dirac mass matrices
having $SU(2)_{L}$-breaking and $SU(2)_{L}$-invariant entries 
respectively, and
${\tilde M}, M$ are Majorana mass matrices 
with $SU(2)_{R} \times U(1)_{G}$-breaking elements for the
mirror and ordinary neutrinos respectively. 
The structure of these matrices  in generation space determines 
the mass hierarchies and mixings of the neutrinos.

Since the present model does not predict the existence of a light
sterile neutrino, contrary to other ``mirror" models \cite{Moha}, 
attention is restricted  to current experimental data
regarding solar and atmospheric neutrino anomalies which imply  
differences of masses-squared 
$\Delta m^{2}_{ij} \equiv m^{2}_{i}-m^{2}_{j}$, 
with $i,j=1,2,3$, and mixing angles $\theta$
between only three mass eigenstates $m_{1,2,3}$
of standard-model left-handed
neutrinos, if one accepts the view that they are due to quantum-mechanically
coherent oscillations between different neutrino-flavour
eigenstates. Assuming for instance that the small-angle MSW solution
to the solar-neutrino deficit involves the left-handed
electron and muon
standard-model neutrinos $\nu_{e}$ and $\nu_{\mu}$, one gets 
experimental bounds which according to \cite{Bilenky} are given by 
\begin{equation}
4 \times 10^{-6} {\rm eV}^{2} \;~^{<}_{\sim} 
\;\Delta m^{2}_{21} \;~^{<}_{\sim} \;1.2 \times 10^{-5} {\rm eV}^{2}
\end{equation}
with a 
mixing angle $\sin{\theta_{\rm sun}} \approx 0.03-0.05$. 

Further information
on neutrino masses and mixing coming from the atmospheric neutrino anomaly
assuming it involves the left-handed standard-model neutrinos
$\nu_{\tau}$ and $\nu_{\mu}$,  gives the bounds 
\begin{equation}
4 \times  10^{-4} {\rm eV}^{2} 
\;~^{<}_{\sim}\; \Delta m^{2}_{31} \;~^{<}_{\sim}
\;8 \times 10^{-3} {\rm eV}^{2}
\end{equation}
with a mixing angle $\sin{\theta_{\rm atm}} \approx 0.49-0.71$
associated with it \cite{Bilenky}. It has to be noted
that this $\nu_{\mu}-\nu_{\tau}$ mixing is unusually large
compared with  charged fermion mixings observed so far.
This experimental input constitutes 
the basis which  determines the form of the neutrino submatrices given
below.

\subsection{Two numerical examples of mass matrices}

The working assumption is made next that the $SU(2)_{L}$-breaking
mirror-fermion Dirac masses satisfy the inequalities 
$m_{U^{M}} > m_{\nu^{M}} > m_{l^{M}}$ for each fermion generation,  
where $U^{M}$ stands for an up-type mirror quark field. This is done
in analogy with the standard model, where each generation contains quarks
which are heavier than the corresponding leptons. 
Up-type quarks are heavier than
the down-type quarks only for the two heavier standard-model 
generations, but it is imagined in the present
example that this is a general property
for mirror quarks and leptons in all their generations. 

Taking into consideration the charged-fermion mass matrices \cite{georg} and
the $SU(2)_{R}$-symmetry breaking scale,
the phenomenological input given above leads to the 
following choice of mass matrices: 
\begin{eqnarray}
\begin{array}{c}  \\ m_{B} ({\rm GeV}) = \\ \end{array} &&
\left(\begin{array}{ccc}
250 & 0 & 0 \\
0 & 350 & 0 \\
0 & 0 & 350
\end{array} \right)
\begin{array}{c}  \\, \; m_{I} ({\rm GeV}) = \\ \end{array}
\left(\begin{array}{ccc}
20 & 1 & 1 \\
1 & 50 & 20 \\
1 & 20 & 70
\end{array} \right)
\end{eqnarray}

\noindent and 
\begin{equation}
{\tilde M} = 0, \; M = 10^{13}\;I_{3} \;\;\; {\rm GeV},  
\end{equation} 
with the $m_{B}$ entries being generated as in the matrix
$M_{L}$ by the strong
$SU(3)_{2G}$ interaction which breaks $SU(2)_{L} \times U(1)_{Y}$
dynamically \cite{georg}, and $I_{3}$ the $3 \times 3$ 
identity matrix. 
Dirac mirror neutrinos are initially considered, since ${\tilde M}=0$. 
It will become clear in the following  that the magnitude of 
the $m_{I}$-entries, in conjunction with $M$,
seems to be crucial in order to
reproduce the correct mixings of Eqs. 16 and 17. Once the magnitude of the  
$M$ entries is fixed,   the structure of $m_{I}$ is therefore
more or less  constrained. 

In the next section  
Majorana mirror neutrinos are also studied, which leads to
the introduction of a ${\tilde M} \neq 0$, 
but it is noted that the Dirac or Majorana nature
of the mirror neutrinos does not influence substantially
the masses and mixings of 
the ordinary neutrinos for which one has experimental evidence to 
compare with the relative theoretical  predictions. 
Therefore, Dirac mirror neutrinos are suitable for the purposes of 
this section.
The scale of the large Majorana masses is taken 
to be close to the scale where the $SU(2)_{R}$ symmetry is broken.
(Gauge invariance dictates of course that this breaking is due to 
a non-zero vacuum expectation value of an
$SU(2)_{R}$-triplet.) 
The matrix $M$ is chosen diagonal for simplicity, even though 
in principle the large 
$\nu_{\mu}-\nu_{\tau}$ mixing could  originate also from this matrix.

Moreover, the mirror neutrino matrices $m_{B}, {\tilde M}$ are also chosen 
diagonal for simplicity, since mirror mixing can only indirectly be fed down
to ordinary fermions and can hardly account for the large 
observed $\nu_{\mu}-\nu_{\tau}$
neutrino mixing. Therefore, even if mirror-neutrino mixing is  
present, lack of relevant experimental evidence would just burden the 
numerical example with more parameters, so it is ignored. 
What is more, large non-diagonal elements in $m_{B}$ would 
lead to dangerously light weak-doublet mirror neutrinos. 
One is therefore left to try large non-diagonal elements for the
matrix $m_{I}$ in order to explain at least part of
the large neutrino mixing. 

The numerical example above leads after diagonalization of the
mass matrix to the following neutrino mass hierarchy
 (at 2 TeV ): 

\noindent Standard-model (Majorana) neutrinos  \hfill Mirror (Dirac)
neutrinos (GeV)$\;$ \\
$m_{\nu_{3L}} =0.03$ eV , $m_{\nu_{2L}}=0.002$ eV,
$m_{\nu_{1L}}=0.00003$ eV \hfill
$m_{\nu_{3^{M}}} = 201$,
$m_{\nu_{2^{M}}} = 200$, \\ 
$m_{\nu_{3R}} \approx m_{\nu_{2R}} \approx   
m_{\nu_{1R}}=10^{13}$ GeV
\hfill $m_{\nu_{1^{M}}} = 180$\\
The mass matrix ${\cal M}_{N}$ gives also a  mixing $\sin{\theta}$ 
for the
solar and atmospheric neutrino problems (which in our case of course
involve only standard-model neutrinos). 
equal to  0.04 and 0.53 respectively, which is
compatible with experiment \cite{Bilenky}.

Whereas the form of $m_{B}$ is consistent with the corresponding
mirror 
charged-lepton matrix $M_{L}$, the matrix $m_{I}$ is slightly problematic,
since it has a gauge-invariant mass term
${\bar \nu_{\mu\;L}}\nu_{\mu\;R}^{M}$ for the second generations  
which is larger
than the corresponding ${\bar c_{L}}c_{R}^{M}$ quark-mass term in \cite{georg}. 
Qualitatively one would generally expect such terms involving 
quarks to be larger than the corresponding lepton ones. Although
this is generally expected 
in analogy with the standard-model case and based to possibly  QCD-related
contributions to particle masses, the lack of a definite calculational 
scheme for these gauge-invariant masses pose limits to such arguments. 
There are nevertheless several solutions to this naturalness issue, and these
are presented below. 

One is to consider lighter mirror leptons, which would then  
allow for smaller entries in $m_{I}$. This by itself would however
not be enough to remove this discrepancy without exceeding  the lower 
mass bounds on direct production of new weak-doublet fermions set by the
LEP experiments. Another 
solution is to consider heavier mirror quarks and charged leptons, 
which would then
require larger entries in the corresponding gauge-invariant quark-mass 
submatrices \cite{georg}. This solution would also help indirectly to
reduce the problems
with the electroweak precision tests, as is shown in the 
next section. 

A third alternative solution to this naturalness question
is here investigated,
i.e. smaller Majorana masses for the fields $\nu_{R}$.  
This is not such a severe assumption, since in nature one has 
already  examples like the electron which has a mass more than five
orders of magnitude smaller than the scale where the symmetry
which forbids its mass breaks. The choice 
$M = 10^{11} I_{3}$ GeV is made next, and  $m_{B}$ is kept the same as before, 
in which case the matrix $m_{I}$ takes 
the form
\begin{equation}
\begin{array}{c}  \\m_{I} ({\rm GeV}) = \\ \end{array}
\left(\begin{array}{ccc}
2 & 1 & 0 \\
1 & 17 & 5 \\
0 & 5 & 21
\end{array} \right).
\label{eq:matr}
\end{equation}

As in the previous case, 
the closeness of the (2,2) and (3,3) entries is crucial if one 
wants to generate large $\nu_{\mu}-\nu_{\tau}$ mixing without 
non-diagonal (2,3) 
and (3,2) entries which would be inconsistently large in comparison
with the corresponding entries of the other fermion mass matrices. Even though
non-diagonal 
entries could also reproduce the correct neutrino mixing by being
 smaller than the ones chosen here, provided
 the diagonal entries were even closer to each-other,
something which would be reminiscent 
of the maximally mixed
$K^{0}-{\bar K^{0}}$ system, such a scenario would fail to 
produce the required mass hierarchies.
Using the mass matrix in Eq.\ref{eq:matr} 
gives rise to the following neutrino mass hierarchy
 (at 2 TeV ): 

\noindent 
Standard-model (Majorana) neutrinos \hfill Mirror (Dirac) neutrinos (GeV)$\;$ \\
$m_{\nu_{3L}} =0.05$ eV , $m_{\nu_{2L}}=0.002$ eV,
$m_{\nu_{1L}}=0.00004$ eV \hfill
$m_{\nu_{3^{M}}} = 201$,
$m_{\nu_{2^{M}}} = 200$, \\ 
$m_{\nu_{3R}} \approx m_{\nu_{2R}} \approx   
m_{\nu_{1R}}=10^{11}$ GeV
\hfill $m_{\nu_{1^{M}}} = 180$.\\
and mixing angles similar to the ones in the previous case.

Even smaller Majorana masses would 
potentially lead to neutrino masses on the order of 1 eV, which 
could be of cosmological interest because of hot dark matter. 
However, having Majorana neutrino masses even lighter than two orders of
magnitude smaller than the $SU(2)_{R}$-breaking scale seems
unlikely and such a possibility is not studied. A similar 
possibility would be to have a  contribution on the order of 1 eV
to the Dirac sector
of all standard-model neutrinos due to unspecified effects, but since
the present model cannot calculate or predict such effects  
this issue is also not pursued further.

It is worth noting here that, in order to get the observed  
fermion mass hierarchies and mixings \cite{georg}, one is lead to consider 
gauge-invariant fermion mass terms ${\bar \psi_{L}^{M}}\psi_{R}$, denoted by
$m_{I\;\psi}$ for $\psi=t,c,b,s,\tau,\mu,\nu_{\tau},\nu_{\mu}$, 
exhibiting  the hierarchy
\begin{equation}
\frac{m_{I\;t}}{m_{I\;c}}, \frac{m_{I\;b}}{m_{I\;s}} \gg 
\frac{m_{I\;\tau}}{m_{I\;\mu}} \gg 
\frac{m_{I\;\nu_{\tau}}}{m_{I\;\nu_{\mu}}}. 
\end{equation} 

\noindent It therefore seems that the more gauge interactions a 
fermion flavour has, the larger the gauge-invariant-mass 
splitting  between its third- and second-generation representatives, even
when  these gauge
interactions  are relatively weak compared to the $SU(3)_{2G}$ one.
This could be an indication 
of  near-critical four-fermion interactions 
in the sense explained in \cite{Miransky} 
and which are contained in  the scenario presented in \cite{georg}. 
The large neutrino mixing suggested by the SuperKamiokande data
is therefore consistent with the existence of such type of interactions, 
as made clear by the generic pattern noted in Eq. 21. 

\subsection{The mixing parameters}

The symmetric charged- and neutral-lepton mass matrices used above
are diagonalized by the 
$6 \times 6$ and $12 \times 12$ matrices which we denote by  
$K_{i}, i=L, N$ respectively, via the relations 
\begin{equation}
{\cal M}_{i} = K_{i}J_{i}K^{\dagger}_{i},
\label{eq:mixing1}
\end{equation}
where the  $J_{i}$ denote  
diagonal matrices.
The lepton mixing information is therefore contained in
a $6 \times 12$ matrix $U$ defined by the relation
\begin{equation}
U_{lj} = (K_{L}^{\dagger})_{lm}(K_{N})_{\nu_{m}j}, 
\label{eq:mixing2}
\end{equation}
where
$m$-summation is implied, with    
$l, m = e, \mu, \tau, e^{M}, \mu^{M}, \tau^{M}$, with $\nu_{m}$
running only over the $SU(2)_{L}$-doublet neutrino flavours,  
 and $j =1, ..., 12$. 
Note that, following the convention of 
\cite{Bilenky} and contrary to the neutrino case, we
keep flavour indices for the charged-lepton mass eigenstates due to the
assumed small mixing between them. 

It is therefore clear that 
the three ordinary-neutrino and three mirror-neutrino
flavor eigenstates  $\nu_{l}$ which are weak doublets are 
 given in terms of the twelve neutrino mass eigenstates $\nu_{j}$
via the relation $\nu_{l} = U_{lj}\nu_{j}$, with $j$-summation implied. 
The first three neutrino mass eigenstates are light enough to allow  
their superposition to be considered as coherent. 
Furthermore, since the matrix $M_{L}$ is almost diagonal, $K_{L}$ is close
to the unit matrix and the form of $U$ is 
mostly affected by $K_{N}$.

Experimentally
one has presently information only on some of the elements of a
$3 \times 3$ submatrix of $U$ involving standard model
left-handed neutrinos and denoted by $U^{SM}_{st}$, with 
$s=e,\mu,\tau$ and $t=1,2,3$. The above
mass matrices allow the calculation of
$U^{SM}$ by means of Eqs.\ref{eq:mixing1} and \ref{eq:mixing2}, 
and this is found to be equal (in absolute values) to  
\begin{equation}
\begin{array}{c}  \\ |U^{SM}| =\\ \end{array} 
\left(\begin{array}{ccc} \sim 1 & 0.039 & 0.01 \\ 0.04 & 0.87 & 0.5 \\
0.008 & 0.5 & 0.86 \end{array} \right).  
\end{equation}

\noindent Its form is quasi-symmetric as expected by the form of
the mass matrices assumed.
This is consistent with the matrix given in \cite{Bilenky} for the small-angle
MSW solution to the solar-neutrino deficit, even though
in the present case $U^{SM}$ is not rigorously unitary because of the 
existence of mirror leptons which slightly mix with the ordinary ones. 
Moreover, it is observed that 
the smallness of the element $U^{SM}_{e3}$ justifies in the 
current example the assumption
that the two oscillations are practically decoupled \cite{Bilenky}. 

Larger non-diagonal entries in the matrix $m_{I}$ can further
increase the entries (2,3) and (3,2) and the 
corresponding $\nu_{\mu}-\nu_{\tau}$ mixing. A similar analysis
therefore could also be easily  performed for the large mixing-angle MSW
and the vacuum-oscillation solutions for the solar-neutrino problem, 
without altering the conclusions drawn above 
about the possibility of having heavier mirror fermions than  previously 
imagined for naturalness reasons, since these are based only on the heavier
neutrino mass eigenstates. 

\section{Mirror neutrinos and the $S$ parameter} 

The contributions of the mirror fermions to the electroweak 
precision parameters $S$ and $T$ were calculated in \cite{georg}
assuming Dirac mirror neutrinos. Since the 
generation symmetries which prohibit mirror Majorana masses 
are broken at around 2 TeV, it is natural to consider Dirac-Majorana
mirror neutrinos next.
This can be achieved  by introducing a non-zero
matrix ${\tilde M}$ with entries near that scale, 
for example ${\tilde M} = 600 I_{3}$ GeV. 
 The standard-model masses and
mixings do not change substantially with this introduction, 
while the  Dirac-Majorana mirror neutrino
mass hierarchy takes now the form (in GeV): 

\noindent $m_{\nu^{M}_{3R}} =169$, $m_{\nu^{M}_{2R}}=162$,
$m_{\nu^{M}_{1R}}=91$  \hfill
 \\ 
$m_{\nu^{M}_{3L}} =768$, $m_{\nu^{M}_{2L}}=761$,
$m_{\nu^{M}_{1L}}=691$.  

By making the identification $m_{a} \equiv m_{\nu^{M}_{R}}$ and
$m_{b} \equiv m_{\nu^{M}_{L}}$ for notational convenience
for each of the three mirror-neutrino generations, in
the limit $m_{a,b,l} \gg m_{Z}$ the ``oblique"
leptonic contribution to the $S$ parameter for 
each mirror generation having a charged lepton of mass $m_{l}$ is given
by \cite{Terning} 
\begin{eqnarray}
S^{0}_{l}&=&\frac{1}{6 \pi}\{c^{2}_{\theta}
\ln{(m_{a}^{2}/m^{2}_{l})}+s_{\theta}^{2}\ln{(m_{b}^{2}/m^{2}_{l})}+
3/2 - \nonumber \\ 
&& 
s^{2}_{\theta}c^{2}_{\theta} [ 8/3+f_{1}(m_{a},m_{b})-
f_{2}(m_{a},m_{b})\ln{(m_{a}^{2}/m^{2}_{b})} ] 
\},  
\end{eqnarray}

\noindent where
\begin{eqnarray}
f_{1}(m_{a},m_{b})&=&\frac{3m_{a}m_{b}^{3}+3m_{a}^{3}m_{b}
-4m^{2}_{a}m_{b}^{2}}{(m_{a}^{2}-m_{b}^{2})^{2}} \nonumber \\ 
f_{2}(m_{a},m_{b})&=&\frac{m_{a}^{6}-3m_{a}^{4}m_{b}^{2}+6m_{a}^{3}m_{b}^{3}-
3m_{a}^{2}m_{b}^{4}+m_{b}^{6}}{(m_{a}^{2}-m_{b}^{2})^{3}}. 
\end{eqnarray}
This result is identical to the one given in \cite{Berndt} in this 
mass-limit only if the
quantities $c_{\theta}$ and $s_{\theta}$ are correctly defined as 
\begin{equation}
c_{\theta}^{2}=1-s_{\theta}^{2} = m_{b}/(m_{a}+m_{b}). 
\end{equation}

In the above, corrections due
to the fact that one mirror neutrino is not much heavier than the $Z$ boson
are neglected, since the purpose of this example is
to just illustrate an effect which depends only 
on mass ratios and not on independent masses, and since
one has poor knowledge of the overall mirror fermion mass
normalization anyway. Note
moreover that, contrary to \cite{Terning}-\cite{Berndt}, 
the mirror neutrino masses $m_{a,b}$ do not
correspond to pure weak eigenstates due to mixing with ordinary 
neutrinos. This mixing is however small due to the relative 
smallness of the elements of $m_{I}$ compared to the $m_{B}$ entries, 
and its effects are therefore also neglected. 

As regards the $\Delta \rho$ parameter which measures
the isospin breaking in the new sector, it
is shown in \cite{georg} that there exists no problem rendering it
small enough to fit experiment, even though some 
fine tuning might be needed. Since whatever leptonic contributions
due to Majorana mirror neutrinos as described
in \cite{Terning}-\cite{Berndt} 
can be compensated by a corresponding shift to the up-down mass
splitting of the mirror fermions, there is no use of further discussing it
in the present context when the precise mirror-fermion mass spectrum remains
experimentally unknown.

The total oblique correction $S^{0}$ 
to $S$ in this model assuming QCD-like dynamics 
is the sum of the contributions $S^{0}_{q}$ and $S^{0}_{l}$ 
coming from mirror quarks and leptons respectively, i.e. \cite{georg}
\begin{equation}
S^{0}=S^{0}_{q}+S^{0}_{l} = 0.9+0.3.
\end{equation}
For the light mirror charged leptons chosen in the
previous section, the change in $S^{0}_{l}$ due to the Dirac-Majorana
nature of the mirror neutrinos is 
marginal, i.e. $S^{0}_{l}=0.12$ instead of $S^{0}_{l}=0.3$ for the
case of Dirac mirror neutrinos. 
If one chooses heavier charged mirror leptons, the change in $S^{0}_{l}$
is larger.  For instance, for 
$m_{l^{M}} \approx 400$ GeV one gets $S^{0}_{l} = -0.12$, and for
$m_{l^{M}} \approx 600$ GeV one gets $S^{0}_{l} = -0.24$.  
The fact that   negative $S$ values are currently 
favored by experiment \cite{Bogdan} could therefore  
 be an indication that the mirror leptons 
are heavier than   the ones of about 200 GeV 
chosen in \cite{georg}. The lightest mirror neutrino cannot be
much lighter than what it is taken here because smaller values for its
mass are excluded by present experiments. 

After analyzing the above formula for $S$, it is concluded that
contributions to $S_{l}$  are not very sensitive to $m_{b}$, but
depend drastically on $m_{a}/m_{l}$. Heavier mirror leptons 
would produce an even smaller $S$ parameter, but assuming that the
mirror quarks are at least as heavy as them would render 
difficult the correct reproduction of the weak scale after a
certain point.
It is nevertheless clear that a larger $m_{a}/m_{l}$ hierarchy could
facilitate the reproduction of a small or even negative $S$ parameter
in accordance with experiment. This could be achieved now with 
the assistance of vertex corrections and non-QCD-like dynamics in these
models as described in \cite{georg} without having to 
introduce very large top-quark anomalous couplings. Such a situation would
also reduce the fine-tuning needed to keep the $\Delta \rho$ parameter small.

\section{Conclusions}

Mirror fermions near the weak scale offer rich possibilities for
the study of new physics. However,
the absence of direct experimental evidence on the existence of 
mirror partners to the standard-model fermions lead  to the
present qualitative
study of various unification possibilities and related neutrino physics
without a prior knowledge of the exact mirror mass hierarchies.
This did not prevent however  very useful general conclusions 
pertinent to such types of models to be drawn. There are two
basic results  to be kept in mind. One is that
unification of all the gauge couplings, including the generation group
coupling, is possible within this group-theoretical 
context and consistent not only with the weak scale but also with
current bounds on the proton life-time. The other is that neutrino masses and
mixings consistent with the observed solar and atmospheric neutrino anomalies
are naturally achieved.

In particular, it is made clear that 
with the proposed fermion content extension, 
not only $SU(5)$ unification is disfavored, but also 
unification with an $SU(3)_{2G}$ generation group is possible.  
Within the group-theoretic framework chosen, this is possible by
taking the gauge coupling ($g_{1}$) of one sector
to be much larger than the other ($g_{2}$) at the 
unification scale. 
This unification is {\it a priori} not at all 
obvious, and constitutes a highly
non-trivial result within the context of dynamical symmetry breaking theories. 
There exist
nevertheless no direct indications  that the scales $\Lambda_{M}$ and
$\Lambda_{PS}$ assume indeed
the exact values needed for this to happen, and no guarantee
that this coupling crossing is not just a coincidence with no particular
importance for the  embedding of the standard-model gauge structure. 

Moreover, the existence of ``mirror fermions" which are weak singlets is 
also not favored. This is a clear manifestation of the 
``$sin{\theta_{W}}$" problem in \cite{Cho}, \cite{Buras} 
which does not appear when weak-doublet  
mirrors of the type introduced in \cite{georg} are used. Even though
these could {\it a priori} pose problems with the $S$ parameter, it
was recently shown \cite{georg} that vertex corrections could alleviate
these effects. 
It is further shown that the most probable symmetry breaking channel is the
$SU(4)_{PS} \times SU(2)_{R} \longrightarrow SU(3)_{C} \times U(1)_{Y}$. 
This breaking channel corresponds to 
a unification scale $\Lambda_{GUT}$ small enough
to suggest that detection of   
proton decay could be soon experimentally accessible. 

Present bounds 
on proton decay further indicate that no more fermion  generations are
very probable at low scales, since then, even though
unification would still be possible, the unification coupling would
be too large.
Furthermore, the different running of the $SU(3)_{C}$ and $SU(3)_{2G}$
gauge couplings due to the  different fermion numbers
which correspond to them provide a 
natural and very interesting explanation of the hierarchy between the
QCD scale and the weak scale, i.e. approximately the scale where the
generation interactions $SU(3)_{2G}$ become strong.

The unification investigation  conducted makes also apparent a problem
in this theory having to do with the generation of lepton masses. In 
particular, the Pati-Salam scale $\Lambda_{PS}$ is found to be 
too large to allow
quark masses to be fed down to leptons via effective four-fermion
operators associated to the $SU(4)_{PS}$ breaking. 
If one does not want to use a fundamental Higgs mechanism 
to break the generation group at the TeV-scale, 
a solution to this problem would be a strong $U(1)_{G}$ at the TeV scale 
\cite{georg}. 
To avoid a Landau pole to the corresponding gauge coupling, the
group $U(1)_{G}$ has to be soon embedded into a larger non-abelian group,
like $SU(4)_{2G}$. 

However,  if one insists on unifying the generation coupling with
the rest of the gauge couplings the solution above is unfortunately
not viable, since as was noted in subsection 2.4 the 
$SU(4)_{2G}$ coupling runs to fast to unify with the other gauge couplings.
The group $SU(4)_{2G}$ has therefore 
to be broken at the unification scale and the $U(1)_{G}$
coupling at low energies is consequently
very weak. A way out for lepton-mass generation
could in principle be the existence of gauge-invariant
operators generated beyond tree-level which feed down quark masses to leptons. 

As already stressed, in the unification analysis 
presented several effects are neglected.
These are related to 
unification threshold effects,  to the Higgs content needed to 
break the $SU(4)_{PS}$ and $SU(2)_{R}$ symmetries, to 2- and higher-loop 
contributions to the $\beta$ functions, to the fact that the
mirror fermions are taken to be all degenerate in mass,  and to the fact that 
$SU(3)_{2G}$ becomes strongly coupled at around 1 TeV, 
something that could influence the rest of the couplings. It is not expected
however that these effects can spoil the  qualitative results of
the analysis above. Unification
could still be achieved if these effects were correctly
taken into account, since
one has the freedom to adjust the scales  
$\Lambda_{M}$ and  $\Lambda_{PS}$ without influencing considerably the
unification scale $\Lambda_{GUT}$. This is particularly true for the
favored possibility presented in Figure 1, since the 
proximity of the scales $\Lambda_{PS}$ and $\Lambda_{GUT}$ does not leave
room for large adjustments. 

One could of course claim that 
the freedom to adjust $\Lambda_{PS}$  in 
order to achieve unification makes this exercise easier to complete and
reduces the predictability of the theory by adding an extra free parameter. 
On the other hand, 
the most favored scenario  described connects this scale 
with the breaking not only of $SU(4)_{PS}$, but also 
 of the $SU(2)_{R}$ symmetry. The examples
involving Majorana neutrinos which are presented above indicate however that
this scale is expected to be several orders of magnitude smaller than 
the unification scale. This not only 
speaks against a ``desert" reaching up to 
$\Lambda_{GUT}$, but is also consistent with the scenario analyzed here.

As already noted, 
the unification considerations above indicate  a favored $SU(2)_{R}$
breaking scale usually associated with the mass of heavy
Majorana neutrinos in the context of the  see-saw mechanism. This leads 
 to the study of 
 neutrino masses and mixings in this framework and in connection
with recent experimental results. 
It is found that, in order to have the  
neutrino masses and mixings compatible with experiment and unification, 
 heavier charged mirror fermions than the ones
quoted in \cite{georg} might be needed, unless the heavy standard-model
Majorana neutrinos
are quite lighter than the scale where $SU(2)_{R}$ breaks. 
Heavier mirror fermions  imply furthermore not only a more difficult 
detection of their indirect effects, since their mixing with
the standard-model fermions is smaller, but also
smaller need of fine-tuning of their masses \cite{georg}. It is further
interesting
to  show that the above observation is perfectly consistent with a
small $S$ parameter which is currently favored by electroweak precision 
tests. A small $S$-parameter could furthermore be an indication that
the lightest mirror neutrino, i.e. the field denoted as $\nu_{1R}^{M}$, 
is so light that it could lie just beyond the reach 
of present high-energy collider experiments. 

\noindent {\bf Acknowledgements} \\
I thank S. Bilenky K. Dick and M. Lindner 
for very helpful discussions. This work is supported by an 
{\it Alexander von Humboldt Fellowship}.

\end{document}